\documentclass[12pt]{article}
\usepackage{amsfonts,amsmath,amssymb,mathrsfs}

\title{Smoothness of Wave Functions in Thermal Equilibrium} 
\author{ 
Roderich Tumulka\footnote{Mathematisches Institut,
    Eberhard-Karls-Unversit\"at, Auf der Morgenstelle 10, 72076
    T\"ubingen, Germany.  E-mail:
    tumulka@everest.mathematik.uni-tuebingen.de}\ \:and
Nino Zangh\`\i\footnote{Dipartimento di Fisica dell'Universit\`a di
    Genova and INFN sezione di Genova, Via Dodecaneso 33, 16146
    Genova, Italy.  E-mail: zanghi@ge.infn.it}
}
\date{September 13, 2005}

\newcommand{\Hilbert}{\mathscr{H}}
\newcommand{\conf}{\mathcal{Q}}
\newcommand{\tr}{\mathrm{tr}}
\newcommand{\Laplace}{\Delta}
\newcommand{\E}{\mathrm{e}} 
\newcommand{\I}{\mathrm{i}} 
\newcommand{\D}{\mathrm{d}} 
\renewcommand{\Re}{\mathrm{Re}}
\renewcommand{\Im}{\mathrm{Im}}
\renewcommand{\sp}[2]{\langle #1 | #2 \rangle}
\newcommand{\EEE}{\mathbb{E}}

\newcommand{\RRR}{\mathbb{R}}
\newcommand{\CCC}{\mathbb{C}}
\newcommand{\ZZZ}{\mathbb{Z}}
\newcommand{\NNN}{\mathbb{N}}
\newcommand{\indexset}{\mathcal{N}}
\newcommand{\prob}{\mathrm{Prob}}
\newcommand{\domain}{\mathrm{domain}}
\newcommand{\sub}{W} 
\newcommand{\sphere}{\mathscr{S}} 
\newtheorem{theorem}{Theorem}
\newtheorem{corollary}{Corollary}
\newenvironment{proof}[1]{\noindent \textit{Proof#1.}\ }
  {\hfill$\square$\bigskip}

\newcommand{\z}[1]{{#1}}

\begin{document}
\maketitle
\begin{abstract}
  We consider the thermal equilibrium distribution at inverse
  temperature $\beta$, or canonical ensemble, of the wave function
  $\Psi$ of a quantum system. Since $L^2$ spaces contain more
  nondifferentiable than differentiable functions, and since the
  thermal equilibrium distribution is very spread-out, one might
  expect that $\Psi$ has probability zero to be differentiable.
  However, we show that for relevant Hamiltonians the contrary is the
  case: with probability one, $\Psi$ is infinitely often
  differentiable \z{and even analytic}.  We also show that with 
  probability one, $\Psi$ lies
  in the domain of the Hamiltonian.

  \medskip\noindent
  MSC (2000):
  \underline{82B10}; 
  60G15; 
  60G17. 
  PACS: 05.30.-d; 
  02.50.-r. 
  Key words: canonical ensemble in quantum physics; Gaussian measures;
  smooth sample paths.
\end{abstract}

\section{Introduction}

We address the question whether the wave function $\Psi$ of a typical
system from the canonical ensemble of thermodynamics with inverse
temperature $\beta$ is differentiable.  As pointed out in
\cite{thermo1}, the thermal equilibrium distribution of the wave
function, corresponding to the canonical ensemble, is the ``Gaussian
adjusted projected measure'' $GAP(\rho)$, a probability measure on the
unit sphere in Hilbert space whose definition we recall in Section~2,
for $\rho = \rho_\beta$, the density matrix of the canonical ensemble,
given by (see, e.g., \cite{LL,tolman})
\begin{equation}\label{rhobetaH}
  \rho_\beta = \frac{1}{Z(\beta,H)} \E^{-\beta H} \quad \text{with }
  Z(\beta,H) = \tr \, \E^{-\beta H}.
\end{equation}
Thus, we take $\Psi$ to be a random unit vector with distribution
$GAP(\rho_\beta)$.  The surprising result is that in many relevant
cases $\Psi$ has probability one to be infinitely often
differentiable \z{and even analytic, i.e.,} $GAP(\rho_\beta) (C^\infty)
= GAP(\rho_\beta) (C^\omega) = 1$.

We explore four kinds of arguments concerning the smoothness of
$\Psi$, each requiring somewhat different assumptions on the
Hamiltonian $H$ and leading to somewhat different conclusions.  Some
of the arguments do not depend on the special measure
$GAP(\rho_\beta)$ but show that, for suitable Hamiltonians $H$, every
distribution whose density matrix is $\rho_\beta$ will be concentrated
on the smooth \z{(resp.\ analytic)} functions; other arguments use the 
way the measure
$GAP(\rho)$ is constructed from a Gaussian measure.  The measure
$GAP(\rho)$ is discussed in detail in \cite{thermo1}. It has density
matrix $\rho$ and is stationary if $\rho$ is.

The first argument aims at showing that the Fourier coefficients of
$\Psi$ go to zero so fast that they are still square-summable after
multiplication by any power of the wave number $k$. This can be easily
applied to cases in which the eigenfunctions of $H$ are plane waves,
such as for the free Schr\"odinger equation in a box.  The second
argument is based on the assumption that the eigenfunctions of $H$ are
smooth, and the theorem asserting that, for a series of functions,
summation and differentiation commute if the series of the derivatives
converges uniformly. We formulate a condition on $H$ that entails this
kind of convergence almost surely (a.s.) for the expansion of $\Psi$
in eigenfunctions of $H$. The third argument, which supposes that
$\Psi$ is a function on an interval $I \subseteq \RRR$, aims at
showing that the increments are not too large, $|\Psi(q + \Delta q) -
\Psi(q)| \lesssim \Delta q$, which suggests differentiability;
however, the rigorous version of this argument provides only a very
weak result. The fourth argument, the simplest and most elegant one,
is of a more abstract nature: it concerns not the question $\Psi \in
C^\infty$ but instead the related question $\Psi \in \domain(H^\ell)$
for $\ell \in \NNN$; indeed, we obtain without further assumptions on
$H$ that a.s.\ $\Psi \in C^\infty(H) := \bigcap_{\ell=1}^\infty
\domain(H^\ell)$ for almost all $\beta$ for which thermal equilibrium
exists at all. \z{We also provide a variant of this argument concerning
the space of analytic vectors of $H$.}

The paper is organized as follows. In Section~2, we recall from
\cite{thermo1} the definition of the measure $GAP(\rho_\beta)$
representing the canonical ensemble. In Section~3, we study
as an example of $H$ the Laplacian on the circle. We conclude
smoothness \z{and analyticity} of $\Psi$ from an analysis of the decay behavior of the
Fourier coefficients of $\Psi$. In Section~4, we apply the same
argument to the (relativistic or nonrelativistic) ideal gas in a box.
In Section~5, we take into account the symmetrization of the wave
function for describing bosons or fermions.  In Section~6, we give the
second kind of argument, providing a general criterion on the
Hamiltonian that is sufficient for concluding that $\Psi$ is a.s.\
smooth. The criterion concerns bounds on the derivatives of the
eigenfunctions of $H$. In Section~7, we discuss the third argument,
which concerns the direct estimation of the difference quotients of
$\Psi$. In Section~8, we describe the fourth argument, which allows to
conclude that $\Psi$ lies in the domain of $H$ and all its powers.  In
Section~9, we conclude, as an application of our results, that $\Psi$
a.s.\ possesses a Bohmian velocity field.

\section{The Canonical Ensemble}

In this section we give the definition of the measure $GAP(\rho)$ on
the unit sphere $\sphere(\Hilbert)$ of Hilbert space $\Hilbert$, as
introduced in \cite{thermo1}.

\z{The} measure $GAP(\rho)$ \z{is defined} for every density matrix $\rho$
(positive operator with $\tr \, \rho =1$) on $\Hilbert$. We obtain the
thermal equilibrium measure $GAP(\rho_\beta)$ by using the canonical
density matrix \eqref{rhobetaH} for a self-adjoint operator $H$ (the
Hamiltonian) and a number $\beta>0$ (the inverse temperature) such
that
\begin{equation}\label{trfinite}
  Z(\beta,H) = \tr \,\E^{-\beta H} < \infty\,.
\end{equation}

The measure $GAP(\rho)$ is defined as the distribution of the random
vector
\begin{equation}\label{Psidef}
  \Psi^{GAP} = \Psi^{GA} / \|\Psi^{GA}\|\,,
\end{equation}
where $\Psi^{GA}$ is a random vector with distribution $GA(\rho)$ (the
``Gaussian adjusted measure'') defined by
\begin{equation}\label{muNdef}
  GA(\rho)(\D\psi) = \|\psi\|^2 \, G(\rho)(\D\psi)\,,
\end{equation}
where $G(\rho)$ is the Gaussian measure on $\Hilbert$ with covariance
matrix $\rho$.

More explicitly, for a random vector $\Psi^G$ to be
$G(\rho)$-distributed means that for any $\phi_1, \phi_2 \in \Hilbert$
the components $Z_1 = \sp{\phi_1} {\Psi^G}$ and $Z_2 = \sp{\phi_2}
{\Psi^G}$ of $\Psi^G$ are complex Gaussian random variables with mean
zero and covariance
\begin{equation}\label{muGcov}
  \EEE Z_1 Z_2^* = \EEE \sp{\phi_1} {\Psi^G} \sp{\Psi^G} {\phi_2} =
  \sp{\phi_1} {\rho| \phi_2},
\end{equation}
where $\EEE$ denotes expectation.  In particular, if $\{
|\varphi_n\rangle \}$ is an orthonormal basis of $\Hilbert$ consisting
of eigenvectors of $\rho$ with eigenvalues $p_n$, then the
coefficients $\sp{\varphi_n}{\Psi^G}$ of $\Psi^G$ are independent
complex Gaussian random variables with mean zero and variances
\begin{equation}\label{Znvar}
  \EEE \bigl| \sp{\varphi_n}{\Psi^G} \bigr|^2 = p_n \,.
\end{equation}
If $\rho$ is of the form \eqref{rhobetaH} then the $\varphi_n$ are
also eigenvectors of $H$.

Although we are interested only in those $\rho$ of the form
\eqref{rhobetaH} for physically relevant Hamiltonians, we will
sometimes, when this makes the mathematics clearer and more elegant,
formulate facts or conditions in terms of an arbitrary density matrix
$\rho$.

For any probability measure $\mu$ on $\sphere(\Hilbert)$, its density
matrix is given by
\begin{equation}\label{rhomu}
  \rho_\mu = \int\limits_{\sphere(\Hilbert)} \mu(\D\psi) \,
  |\psi \rangle \langle \psi |\,,
\end{equation}
or $\rho_\mu = \EEE_\mu |\psi\rangle \langle \psi|$, where $\EEE_\mu$
denotes the expectation with respect to $\mu$. In particular,
$\EEE_\mu \bigl| \sp{\varphi}{\psi} \bigr|^2 = \sp{\varphi}{\rho_\mu|
  \varphi}$ for every fixed $\varphi \in \Hilbert$. (If a probability
measure $\mu$ on $\Hilbert$ is not concentrated on
$\sphere(\Hilbert)$, the notion of density matrix of $\mu$ does not
make sense any more; however, \eqref{rhomu}, or $\EEE_\mu |\psi
\rangle \langle \psi|$, is still the \emph{covariance matrix} of
$\mu$.)  As mathematically expressed by $\EEE_\mu \sp{\psi}{P|\psi} =
\tr(\rho_\mu P)$ for every projection $P$, $\rho_\mu$ provides the
distribution of outcomes of any quantum experiment on a system with
$\mu$-distributed random wave function.  The density matrix of
$GAP(\rho)$ is indeed $\rho$.  This and other fundamental properties of
the measure $GAP(\rho)$ are discussed in \cite{thermo1}.

The following simple fact will sometimes be useful as it reduces the
task of showing smoothness of $\Psi = \Psi^{GAP}$ to showing
smoothness of the Gaussian random vector $\Psi^G$ with distribution
$G(\rho)$.  For any subspace $\sub$ of $\Hilbert$, we have that
\begin{equation}\label{muG1mu1}
  \text{if } G(\rho)(\sub) =1 \text{ then } GAP(\rho)(\sub)=1 \,.
\end{equation}
To see this, note that $GA(\rho)$ has the same null sets as $G(\rho)$
(as it is absolutely continuous with respect to $G(\rho)$ with a
density that vanishes only at one point), so that, if $G(\rho)(\sub)
=1$, $0 = G(\rho)(\Hilbert \setminus \sub) = GA(\rho)(\Hilbert \setminus
\sub)$ and thus $GA(\rho)(\sub)=1$; but, by definition \eqref{Psidef},
if $\Psi^{GA} \in \sub$ then also $\Psi^{GAP} \in \sub$.

\section{A Case Study: The Laplacian on the Circle}

In this section we consider a single particle moving on a circle
$S^1$ with the free Schr\"odinger Hamiltonian,
\begin{equation}\label{Hcircle}
  H = -\frac{\hbar^2}{2m} \frac{\partial^2}{\partial q^2},
\end{equation}
where $m$ denotes the mass of the particle, $q$ the angular
coordinate on the circle, \z{and wave functions are written as periodic 
functions of} $q$. The result we derive is that relative to any
measure $\mu$ on $\sphere(\Hilbert)$ with density matrix $\rho_\beta$
with $\beta>0$ (or, in fact, any measure $\mu$ on $\Hilbert$ with
covariance matrix $\rho_\beta$), almost every wave function $\psi$ is
smooth, $\psi \in C^\infty(\RRR)$; \z{we then go on to show that 
$\mu$-almost every wave function is analytic,} $\psi \in C^\omega(\RRR)$.

We begin with considering, instead of differentiability, a closely
related property: existence (in $L^2$) of the distributional
derivative.  In other words, we consider the property of a
wave function $\psi$ that $|k| \, \widehat{\psi}(k)$ is still square
integrable where $\widehat{\psi}$ is the Fourier transform of
$\psi$. Since the functions we are considering are 
\z{$2\pi$-periodic in $q$, the appropriate} 
property is that the Fourier coefficients $c_k$, defined by
\begin{equation}\label{Fourierseries}
  \psi(q) = \sum_{k \in \ZZZ} c_k \, \E^{\I kq},
\end{equation}
are still square-summable after multiplication with $|k|$.  Let
$\sub^\ell$ denote the $\ell$-th Sobolev space, i.e., the subspace of
$L^2([0,2\pi])$ containing those functions whose Fourier coefficients 
$c_k$ satisfy
\begin{equation}
  |k|^\ell \, c_k \text{ is square-summable, i.e., } \sum_{k\in \ZZZ}
  |k|^{2\ell} \, |c_k|^2 < \infty.
\end{equation}
We ask whether $\psi \in \sub^\ell$ \z{for a random wave function 
$\psi$ with distribution $\mu$}.  Since the eigenfunctions of
$H$ are the plane waves,
\begin{equation}\label{eigenfct1}
  \varphi_n(q) = \frac{1}{\sqrt{2\pi}} \, \E^{\I nq}\,, \quad n \in
  \ZZZ,
\end{equation}
the energy coefficients of a wave function are just the Fourier
coefficients. The eigenvalues are
\begin{equation}\label{HcircleEn}
  E_n = \frac{\hbar^2}{2m} n^2, \quad n \in \ZZZ.
\end{equation}
Thus, our question about $\psi$ amounts to asking for which $\ell
\in \NNN$ we have
\begin{equation}\label{inftyFouriercond}
  \sum_{n \in \ZZZ} n^{2\ell} \bigl|
  \sp{\varphi_n}{\psi} \bigr|^2 < \infty.
\end{equation}
This indeed holds $\mu$-a.s.\ for all $\ell \in \NNN$; to see this,
note that\footnote{Numbers on top of equality signs indicate which
  equation is being applied.}
\begin{subequations}\label{expect}
\begin{align}
  &\EEE_\mu \sum_{n \in \ZZZ} n^{2\ell} \bigl| \sp{\varphi_n}{\psi}
  \bigr|^2 = \sum_{n \in \ZZZ} n^{2\ell} \EEE_\mu \bigl|
  \sp{\varphi_n}{\psi} \bigr|^2 =\\ &\stackrel{\eqref{Znvar}}{=}
  \sum_{n \in \ZZZ} n^{2\ell} \frac{\E^{-\beta E_n}} {Z(\beta,H)}
  \stackrel{\eqref{HcircleEn}}{=} \frac{1}{Z(\beta,H)} \sum_{n \in
  \ZZZ} n^{2\ell} \E^{-(\beta \hbar^2/2m) n^2 } < \infty
\end{align}
\end{subequations}
because for any constant $\gamma>0$ and for sufficiently large $|n|$,
\begin{equation}\label{summable}
  (2\ell+2) \log |n| < \gamma |n|^2  \quad \text{and thus} \quad
  |n|^{2\ell} \, \E^{-\gamma |n|^2} < \frac{1}{|n|^2}.
\end{equation}
If the expectation \eqref{expect} of a $[0,\infty]$-valued random
variable is finite, the variable is a.s.\ finite. Thus,
\begin{equation}
  \mu \Bigl( \sum_{n \in \ZZZ} n^{2\ell} \bigl|
  \sp{\varphi_n}{\psi} \bigr|^2 = \infty \Bigr) = 0 \text{ or }
  \mu(\sub^\ell) = 1
\end{equation}
for all $\ell \in \NNN$.

We now make the connection between $\sub^\ell$ and $C^\ell$, i.e.,
with \z{classical} differentiability: by the Sobolev lemma
\cite[p.~52]{ReedSimon2}, every function in $\sub^\ell$ is equal
Lebesgue-almost-everywhere to a function in $C^{\ell-1}$. Hence, every
function in $\bigcap_{\ell=1}^\infty \sub^\ell$ is equal
Lebesgue-almost-everywhere to a function in $C^\infty$. In particular,
$\mu$-a.s.\ there is a $\phi \in C^\infty$ such that $\psi(q) =
\phi(q)$ Lebesgue-almost-everywhere.

\z{We now turn to analyticity. By a similar argument as given in
\eqref{expect} and \eqref{summable}, one can see that}
\begin{equation}
  \mu \Bigl( \sum_{n \in \ZZZ} \E^{2\alpha |n|} \bigl|
  \sp{\varphi_n}{\psi} \bigr|^2 = \infty \Bigr) = 0
\end{equation}
\z{for every $\alpha>0$, so that}
\begin{equation}\label{expcondition}
  \E^{\alpha |k|} \, c_k \text{ is a.s. square-summable.}
\end{equation}
\z{Regarding the variable $q$ in \eqref{Fourierseries} as complex, 
we observe that the right hand side of \eqref{Fourierseries} converges,
as a consequence of \eqref{expcondition}, uniformly in every strip
$-\alpha +\varepsilon < \Im\, q <\alpha -\varepsilon$ with $0<\varepsilon 
< \alpha$. (To see this, we use that square-summable sequences are bounded, 
$\E^{\alpha |k|} \, |c_k| \leq C$, so that for $q$ in the strip, $|c_k \, 
\E^{\I kq}| = |c_k| \, \E^{-k\, \Im\, q} < |c_k| \, \E^{|k| (\alpha -
\varepsilon)} \leq C \, \E^{-|k|\varepsilon}$, which is summable over 
$k \in \ZZZ$.) Since the uniform limit of analytic functions on an open
set in the complex plane is analytic (by virtue of the Cauchy integral 
formula), $\psi$ is analytic in the strip $-\alpha < 
\Im \, q < \alpha$; since $\alpha$ was arbitrary, $\psi$
is entire (i.e., analytic on the whole complex plane). More precisely, 
$\mu$-a.s.\ there is an entire function $\phi$ such that $\psi(q)
= \phi(q)$ Lebesgue-almost-everywhere in $\RRR$.}

\section{The Ideal Gas in a Box}

In a similar way, we can treat any Hamiltonian whose eigenfunctions
are plane waves. A particularly relevant case is that of the ideal
gas: $N$ noninteracting particles in a $d$-dimensional box
$[0,\pi]^d$ with \z{Hamiltonian}
\begin{equation}\label{Hbox}
  H = -\sum_{i=1}^N \frac{\hbar^2}{2m} \Laplace_i
\end{equation}
\z{with Dirichlet boundary conditions,}
where $\Laplace_i$ is the Laplacian acting on the coordinates of the
$i$-th particle. Our conclusion will again be that $\mu( C^\infty ) = \mu(C^\omega) =
1$ for every measure $\mu$ on $\Hilbert$ whose covariance matrix is
$\rho_\beta$, and in particular for $\mu = GAP(\rho_\beta)$. For the
moment, we ignore the symmetrization postulate; we will treat bosons
and fermions in Section~5.

The Hamiltonian $H$ \z{on} $\Hilbert = L^2([0,\pi]^{Nd})$ 
\z{has} eigenfunctions \cite[p.~78]{box}
\begin{equation}\label{eigenfct2}
  \varphi_n(q) = \Bigl( \frac{2}{\pi}\Bigr)^{Nd/2}
  \prod_{i=1}^N \prod_{a=1}^d \sin(  n_{i,a} q_{i,a})
\end{equation}
where $n = (n_{1,1} , \ldots, n_{N,d}) \in \NNN^{Nd}$ and $q =
(q_{1,1}, \ldots, q_{N,d}) \in [0,\pi]^{Nd}$, and eigenvalues
\begin{equation}\label{eigenval2}
  E_n = \sum_{i=1}^N \sum_{a=1}^d \frac{\hbar^2}{2m} n_{i,a}^2 =
  \frac{\hbar^2}{2m} \|n\|^2.
\end{equation}

The right hand side of \eqref{eigenfct2} extends in an obvious way to
a function on $\RRR^{Nd}$ that is $2\pi$-periodic in every variable,
which we also call $\varphi_n$; using the coefficients
$\sp{\varphi_n}{\psi}$ of the energy expansion we have a natural
extension of any $\psi \in \Hilbert$ to a function $\phi$ on
$\RRR^{Nd}$ that is $2\pi$-periodic in every variable, $\phi = \sum_n
\sp{\varphi_n}{\psi} \, \varphi_n$. Its Fourier coefficients are
\begin{equation}
  c_k = (-i)^{Nd} \Bigl(\prod_{i=1}^N \prod_{a=1}^d
  \mathrm{sign}(k_{i,a}) \Bigr) \Bigl\langle \varphi_{|k_{1,1}|, \ldots,
  |k_{N,d}|} \Big| \psi \Bigr\rangle
\end{equation}
where $k=(k_{1,1}, \ldots, k_{N,d}) \in \ZZZ^{Nd}$ and we set
$\mathrm{sign}(0) =0$. 

\z{We begin with the existence in $L^2$ of the $\ell$-fold distributional 
derivative. We assert that $\mu$-a.s.\ for all $\ell \in \NNN$} 
\begin{equation}\label{sobolev}
  \sum_{k \in \ZZZ^{Nd}} \|k\|^{2\ell} \, |c_k|^2 < \infty.
\end{equation}
This follows from
\begin{subequations}\label{Eestimate}
\begin{align}
  &\EEE_\mu \sum_{k \in \ZZZ^{Nd}} \|k\|^{2\ell} \, |c_k|^2 = 2^{Nd}
  \sum_{n \in \NNN^{Nd}} \|n\|^{2\ell} \, \EEE_\mu \bigl|
  \sp{\varphi_n}{\psi} \bigr|^2 =\\ &\stackrel{\eqref{Znvar}}{=}
  \frac{ 2^{Nd} } {Z(\beta,H)} \sum_{n\in \NNN^{Nd}} \|n\|^{2\ell} \,
  \E^{-\beta E_n} \stackrel{\eqref{eigenval2}}{=} \frac{2^{Nd}}
  {Z(\beta,H)} \sum_{n\in \NNN^{Nd}} \|n\|^{2\ell} \, \E^{-(\beta
  \hbar^2/2m)\|n\|^2} \leq \\ & \stackrel{\eqref{ineq}}{\leq}
  \frac{2^{Nd}} {Z(\beta,H)} \sum_{n\in \NNN^{Nd}} (1+n_{1,1})^{2\ell}
  \cdots (1+n_{N,d})^{2\ell} \, \E^{-(\beta \hbar^2/2m)\|n\|^2}=\\ &=
  \frac{1} {Z(\beta,H)} \Bigl(2\sum_{\nu \in\NNN} (1+\nu)^{2\ell} \,
  \E^{-(\beta\hbar^2/2m) \nu^2} \Bigr)^{Nd}
  \stackrel{\eqref{summable}}{<} \infty,
\end{align}
\end{subequations}
where we used
\begin{equation}\label{ineq}
  \|n\|^2 = \sum_{i,a} n_{i,a}^2 \leq \prod_{i,a} ( 1 + 2n_{i,a} +
  n_{i,a}^2) = \prod_{i,a} (1+ n_{i,a})^2.
\end{equation}

Inequality \eqref{sobolev} means that $\phi$ lies in the Sobolev space
$W^\ell$, and by the Sobolev lemma \cite[p.~52]{ReedSimon2} also in
$C^{m}$ for all $m< \ell-Nd/2$. Since $\ell$ was arbitrary, $\phi$ is
$\mu$-a.s.\ smooth, and thus so is $\psi$, its restriction to
$[0,\pi]^{Nd}$.

The same argument can be applied to the relativistic case, in which
the Hamiltonian is the free Dirac operator
\begin{equation}\label{dirac}
  H= -\sum_{i=1}^N \bigl(\I c \hbar \alpha_i \cdot \nabla_i + m c^2
  \beta_i \bigr)
\end{equation}
with $c$ the speed of light, $m$ the mass, and $\alpha_i$ and
$\beta_i$ the Dirac alpha and beta matrices acting on the $i$-th spin
index of the wave function. Again, one obtains that $\mu(C^\infty) =1$
for all $\mu$ with covariance matrix $\rho_\beta$.

\z{We turn to analyticity and to this end assert that $\mu$-a.s. for 
all $\alpha>0$}
\begin{equation}\label{expdecay}
  \sum_{k \in \ZZZ^{Nd}} \E^{2\alpha \|k\|} \, |c_k|^2 < \infty.
\end{equation}
\z{This follows from the fact that, by the same reasoning as in 
\eqref{Eestimate},} 
\[
  \EEE_\mu \sum_{k \in \ZZZ^{Nd}} \E^{2\alpha 
  \|k\|} \, |c_k|^2 = \frac{2^{Nd}}
  {Z(\beta,H)} \sum_{n\in \NNN^{Nd}} \E^{2 \alpha \|n\|} \, \E^{-(\beta
  \hbar^2/2m)\|n\|^2} < \infty
\]
\z{because for any constant $\gamma >0$ and for all but finitely many
$n \in \NNN^{Nd}$, $2\alpha \|n\| - \gamma \|n\|^2 < -(\gamma/2) \|n\|^2$,
while $\E^{-(\gamma/2) \|n\|^2}$ is summable over $\NNN^{Nd}$ by 
\eqref{summable}. By the same argument as in the last paragraph of 
Section 3, one can conclude from \eqref{expdecay} that $\phi$ is analytic 
in the cylinder $\{q \in \CCC^{Nd}: \| \Im \, q \| < \alpha\}$. Since
$\alpha$ was arbitrary, $\phi$ is entire. Thus, $\mu$-a.s.\ there is an
entire function $\phi$ such that $\psi(q) = \phi(q)$ Lebesgue-almost-everywhere
in $[0,\pi]^{Nd}$. (For the Dirac equation, since the energy eigenvalues grow
like $c\hbar\|k\|$, $\psi$ a.s.\ possesses an analytic continuation to the
cylinder $\|\Im \, q\| < \beta c \hbar/2$.)}

\section{Bosons and Fermions}

In the previous section, we ignored the symmetrization of the wave
function for systems of bosons or fermions. If one takes the
symmetrization into account, one reaches the same conclusion:
smoothness is almost sure. But instead of going through the
calculation of the previous section again, we provide a simple
argument why $GAP(\rho_\beta)$ must be concentrated on $C^\infty$ for
indistinguishable particles (with symmetrized wave functions) if it is
concentrated on $C^\infty$ for distinguishable particles (with
unsymmetrized wave functions).

The symmetric (respectively anti-symmetric) state vectors form a
subspace of $\Hilbert = L^2 ([0,\pi]^{Nd})$; let $P$ denote the
projection to that subspace; the subspace can be written $P\Hilbert$.
Since the Hamiltonian \eqref{Hbox} is invariant under permutations, we
have that $HP = PH = PHP$.  Thus, the canonical density matrix for
indistinguishable particles is
\begin{equation}
  \rho_\beta (PH,P\Hilbert) = c\, P \rho_\beta(H,\Hilbert) P
\end{equation}
where we have made explicit the dependence of $\rho_\beta$ on the
given Hamiltonian and Hilbert space, \z{and $c = Z(\beta,H)/Z(\beta,PHP)$}.

Now observe that for a Gaussian measure with covariance $\rho$, we
have that $G(P\rho P) = G(\rho) \circ P^{-1}$, where $P^{-1}$ is
understood as mapping subsets of $P\Hilbert$ to their pre-images in
$\Hilbert$, in other words $\Psi^{G(P\rho P)} = P \Psi^{G(\rho)}$ in
distribution.

In our case, $P$ is the symmetrization (respectively
anti-symmetrization) operator, which maps smooth functions to smooth
functions. Since $\Psi^G$ is a.s.\ smooth (by the result of the
previous section), so is $P \Psi^G$; with \eqref{muG1mu1} we conclude
that $GAP(\rho_\beta (PH,P\Hilbert)) (C^\infty)=1$. \z{The same argument
works with analyticity.}

\section{A General Sufficient Condition for Smoothness} 

We now present a second kind of argument, different from the one used
in the previous sections; it applies to the measure $GAP(\rho)$ but not
to all measures with density matrix $\rho$. The argument provides us
with a condition, see \eqref{Cinfty} below, on any given density
matrix $\rho$ (and thus, for $\rho = \rho_\beta$, on the Hamiltonian)
ensuring that $GAP(\rho)(C^\infty)=1$.

\begin{theorem}\label{sta:openset}
  Let $\conf$ be an open subset of $\RRR^d$. Suppose
  that the density matrix $\rho$ on $\Hilbert = L^2(\conf,\CCC^m)$ has
  $C^\infty$ eigenfunctions $\varphi_n(q)$ with $\|\varphi_n\| =1$ and
  eigenvalues $p_n$, such that for all $n$ and all $\ell = 0,1,2,3,
  \ldots$, the $\ell$-th derivative of $\varphi_n$ is bounded,
  \[
    \|\nabla^\ell \varphi_n\|_\infty = \sup_{q \in \conf} |\nabla^\ell
    \varphi_n(q)| < \infty,
  \]
  where by absolute values we mean
  \[
    |\nabla^\ell \psi(q)|^2 = \sum_{i_1, \ldots, i_\ell =1}^d
    \sum_{s=1}^m \Bigl| \frac{\partial^\ell \psi_s} {\partial
    q_{i_1} \cdots \partial q_{i_\ell}} (q) \Bigr|^2.
  \]
  If for all $\ell = 0,1,2,3,\ldots$,
  \begin{equation}\label{Cinfty}
    \sum_n \|\nabla^\ell \varphi_n\|_\infty \, \sqrt{p_n} < \infty 
  \end{equation}
  then $GAP(\rho) \bigl(C^\infty(\conf,\CCC^m) \bigr) =1$.
\end{theorem}

\begin{proof}{}
To begin with, for a complex Gaussian random variable $Z$ with $\EEE Z
=0$ and $\EEE |Z|^2 = \sigma^2$ one can determine that
\begin{equation}\label{EZabs}
  \EEE |Z| = \int_{\RRR^2} dx \, dy \, \frac{\sqrt{x^2 +y^2}} {\pi
  \sigma^2} \exp\Bigl(- \frac{x^2 + y^2}{\sigma^2} \Bigr) =
  \frac{\sqrt{\pi}}{2} \sigma.
\end{equation}
Setting $Z=\sp{\varphi_n}{\Psi^G}$, we obtain that
\begin{align}
  \nonumber &\EEE \sum_n \Bigl\|\nabla^\ell \varphi_n \, \sp
  {\varphi_n} {\Psi^G}\Bigr\|_\infty = \EEE \sum_n \bigl| \sp
  {\varphi_n} {\Psi^G} \bigr| \, \|\nabla^\ell \varphi_n \|_\infty =
  \\ \nonumber &= \sum_n \EEE \bigl|\sp{\varphi_n}{\Psi^G} \bigr| \,
  \|\nabla^\ell \varphi_n\|_\infty \stackrel{\eqref{EZabs}}{=} \sum_n
  \frac{\sqrt{\pi}}{2} \sqrt{p_n} \, \|\nabla^\ell \varphi_n\|_\infty
  \stackrel{\eqref{Cinfty}}{<} \infty,
\end{align}
and therefore
\[
  \prob \Bigl( \sum_n \Bigl\|\nabla^\ell \varphi_n \,
  \sp{\varphi_n}{\Psi^G}\Bigr\|_\infty < \infty \Bigr)=1.
\]
Since this is true of every $\ell$, we have that in the expansion
\begin{equation}\label{expansion}
  \Psi^G(q) = \sum_n \varphi_n(q) \, \sp{\varphi_n}{\Psi^G}
\end{equation}
(having $C^\infty$ partial sums), a.s.\ the $\ell$-th derivatives of
the partial sums converge uniformly; and in particular,
\eqref{expansion} itself converges uniformly. It is a standard theorem
(see, e.g., \cite[p.~118]{analysis}) that if a sequence $f_n$ of
functions converges pointwise and the derivatives $\nabla f_n$
uniformly, then the limit function $f$ is differentiable and has
derivative $\nabla f = \lim \nabla f_n$.  Therefore, a.s.\ $\Psi^G \in
C^\infty(\conf)$, and the derivatives are
\begin{equation}
  \nabla^\ell \Psi^G(q) = \sum_n \nabla^\ell \varphi_n(q) \,
  \sp{\varphi_n}{\Psi^G}.
\end{equation}
By \eqref{muG1mu1}, a.s.\ $\Psi \in C^\infty(\conf)$, which completes
the proof.
\end{proof}

By applying this proof to local coordinates, we can generalize the
result to Riemannian manifolds and vector bundles as follows.
\textit{ Let $\conf$ be a Riemannian $C^\infty$ manifold, $E$ a
  $C^\infty$ complex vector bundle over $\conf$ with \z{positive-definite} $C^\infty$
  Hermitian inner products on the fiber spaces, and let $\nabla$ be
  the covariant derivative operator corresponding to a $C^\infty$
  connection on $E$.  Let $\Hilbert = L^2(E)$ be the Hilbert space of
  square-integrable (with respect to the Riemannian volume)
  measurable cross-sections of $E$, and $C^\infty(E)$ the space of smooth
  cross-sections. Suppose that the density matrix
  $\rho$ on $\Hilbert$ has $C^\infty$ eigen-cross-sections
  $\varphi_n(q)$ with $\|\varphi_n\| =1$ and eigenvalues $p_n$, such
  that for all $n$ and all $\ell=0,1,2,3,\ldots$, the $\ell$-th
  covariant derivative of $\varphi_n$ is bounded,
  \begin{equation}
    \|\nabla^\ell \varphi_n\|_\infty = \sup_{q \in \conf} |\nabla^\ell
    \varphi_n(q)| < \infty,
  \end{equation}
  where the absolute values are taken with respect to the Riemannian
  inner product on tangent spaces and the Hermitian inner product on
  fiber spaces.  If for all $\ell = 0,1,2,3,\ldots$,
  \begin{equation}\label{manifoldCinfty}
    \sum_n \|\nabla^\ell \varphi_n\|_\infty \, \sqrt{p_n} < \infty
  \end{equation}
  then $GAP(\rho) \bigl( C^\infty(E) \bigr) =1$.}

\bigskip

\z{Another easy generalization of Theorem~\ref{sta:openset} concerns 
analyticity:   \textit{Let $\conf_\CCC$ be an open subset of $\CCC^d$ and 
  $\conf := \{(q_1, \ldots, q_d) \in \conf_\CCC : q_1, \ldots, q_d \in \RRR\} 
  \subseteq \RRR^d$. Suppose
  that the density matrix $\rho$ on $\Hilbert = L^2(\conf,\CCC^m)$ has
  eigenvalues $p_n$ with normalized eigenvectors $\varphi_n \in 
  C^\omega(\conf_\CCC)$, with $C^\omega(\conf_\CCC)$ the space of 
  $L^2$ functions on $\conf$ that 
  possess analytic continuations to $\conf_\CCC$. If
  for every compact set $K \subseteq \conf_\CCC$,
  \begin{equation}\label{Comega}
    \sum_n \bigl\|\varphi_n |_K \bigr\|_\infty \, \sqrt{p_n} < \infty\,,
  \end{equation}
  (writing also $\varphi_n$ for the analytic continuation and $\varphi_n|_K$ 
  for its restriction to $K$),
  then $GAP(\rho)\bigl( C^\omega(\conf_\CCC) \bigr)=1$.}}

\bigskip

\begin{proof}{}
\z{By \eqref{Comega} and \eqref{EZabs},}
\[
  \EEE \sum_n \bigl\| \varphi_n|_K \bigr\|_\infty \, \bigl| \sp
  {\varphi_n}{\Psi^G} \bigr| = 
  \sum_n \bigl\| \varphi_n|_K 
  \bigr\|_\infty \, \frac{\sqrt{\pi}}2 \sqrt{p_n} < \infty \,,
\]
\z{and thus a.s.\ $\sum_n \bigl\| \varphi_n|_K \bigr\|_\infty \, \bigl| \sp
{\varphi_n}{\Psi^G} \bigr| < \infty$. As a consequence, the expansion}
\begin{equation}
  \Psi^G(q) = \sum_n \varphi_n(q) \, \sp{\varphi_n} {\Psi^G}
\end{equation}
\z{converges not only for $q \in \conf$ but also for $q \in \conf_\CCC$, 
in fact uniformly on every compact set $K \subseteq \conf_\CCC$.
Since uniform limits of analytic functions are analytic, $\Psi^G$ 
and thus also $\Psi^{GAP}$ are a.s.\ analytic.}
\end{proof}

In order to demonstrate that the conditions \eqref{Cinfty},
\eqref{manifoldCinfty} \z{and \eqref{Comega}} are not 
unreasonably strong, we show that they are satisfied for 
the Laplacian on the circle. Here the eigenfunctions
are given by \eqref{eigenfct1}, and their derivatives, 
\z{respectively analytic continuations to disks $K = \{ q \in 
\conf_\CCC=\CCC: |q| \leq \alpha \}$}, have bounds
\[
  \|\nabla^\ell \varphi_n\|_\infty = \frac{|n|^\ell}{\sqrt{2\pi}}, \quad
  \bigl\|\varphi_n|_K \bigr\|_\infty = \frac{1}{\sqrt{2\pi}} \E^{\alpha |n|}  \,.
\]
In this case, \eqref{Cinfty} respectively \eqref{manifoldCinfty}
is satisfied since
\[
  \sum_{n\in \ZZZ} \|\nabla^\ell \varphi_n\|_\infty \, \E^{-\frac12
  \beta E_n} = \frac{1}{\sqrt{2\pi}} \sum_{n\in \ZZZ} |n|^\ell \,
  \E^{-(\beta \hbar^2/4m) n^2} < \infty
\]
by \eqref{summable}, \z{and similarly \eqref{Comega} since}
\[
  \sum_{n \in \ZZZ} \bigl\| \varphi_n|_K \bigr\|_\infty \, \E^{-\frac12
  \beta E_n} = \frac{1}{\sqrt{2\pi}} \sum_{n \in \ZZZ} \E^{\alpha |n|}
  \, \E^{-(\beta \hbar^2/4m) n^2} < \infty \,.
\]

\section{Estimating Difference Quotients}\label{sec:increments}

In this section, we follow another line of reasoning for studying
regularity properties of $\Psi$, based on a standard theorem on the
regularity of sample paths of a Gaussian process. However, the result
is much weaker than what we obtained in the previous section.

Assume for simplicity that the configuration space is an open interval
$I \subseteq \RRR$, and that $\Hilbert = L^2(I,\CCC)$.  The idea in
this section is to consider the increments $\Psi^G(q+\Delta q) -
\Psi^G(q)$ of the Gaussian process $\Psi^G$ and to argue that for
reasonable Hamiltonians they are of the order of magnitude of $\Delta
q>0$,
\begin{equation}\label{orderh}
  \bigl| \Psi^G(q+\Delta q) - \Psi^G(q)\bigr| \lesssim \Delta q,
\end{equation}
which suggests that difference quotients converge to differential
quotients as $\Delta q \to 0$, i.e., that $\Psi^G$ be
differentiable. However, what can rigorously be concluded from a
statement about the variance of the increment analogous to
\eqref{orderh} is less than differentiability, namely H\"older
continuity with exponent $1-\varepsilon$.

We now describe the argument in detail. We pretend that $\Psi^G$ is
everywhere defined in $I$ (although, strictly speaking, vectors in
Hilbert space are equivalence classes of functions coinciding
Lebesgue-almost-everywhere) in such a way that it is a Gaussian
process in the sense that, for any choice of $q_1, \ldots, q_n \in I$,
the joint distribution of $\Psi^G(q_1), \ldots, \Psi^G(q_n)$ in
$\CCC^n$ is Gaussian. It follows that for any $\Delta q>0$, the
increment $\Psi^G(q+\Delta q) - \Psi^G(q)$ is a Gaussian variable, and
we can compute its variance,
\begin{equation}\label{varincrement}
  \EEE \bigl| \Psi^G(q+\Delta q) - \Psi^G(q)\bigr|^2 = \rho(q+\Delta
  q,q+\Delta q) - 2\Re \,\rho(q,q+\Delta q) + \rho(q,q),
\end{equation}
where $\rho(q,q') = \sp{q}{\rho|q'}$ are the ``matrix elements'' in
the position representation of the density matrix $\rho = \rho_\beta$.
Assuming that
\begin{equation}\label{rhosmooth}
  \rho(q,q') \text{ is a smooth function,}
\end{equation}
which would appear to be a reasonable assumption on the Hamiltonian,
we can employ a Taylor expansion of $\rho$ to the second order around
$(q,q)$ and obtain from \eqref{varincrement} that
\begin{equation}\label{varincrement2}
  \EEE \bigl| \Psi^G(q+\Delta q) - \Psi^G(q)\bigr|^2 =
  \frac{\partial^2 \rho} {\partial q \, \partial q'} \bigg|_{q'=q} \,
  \Delta q^2 + O(\Delta q^3).
\end{equation}
It is a standard result \cite[Thm.~8 of Chap.~III]{GS} that for a
Gaussian process $\Psi^G$ with the following bound on the variances of
the increments:
\begin{equation}\label{varbound}
  \EEE \bigl| \Psi^G(q+\Delta q) - \Psi^G(q)\bigr|^2 \leq K \, \Delta
  q^p,
\end{equation}
where $K>0$ and $p>0$ are constants, the realization a.s.\ satisfies
\begin{equation}\label{GSresult}
   \bigl| \Psi^G(q+\Delta q) - \Psi^G(q)\bigr| \leq K' \, \Delta
   q^{p/2} \, \bigl|\log \Delta q \bigr|^{1+\delta}
\end{equation}
for arbitrary $\delta>0$ and a suitable constant $K' = K'(\delta)>0$.
Inserting \eqref{varincrement2} into \eqref{varbound}, we obtain from
\eqref{GSresult}, in case $\partial^2 \rho/\partial q \, \partial q'
\neq 0$, that a.s.\
\begin{equation}\label{asbound}
  \bigl| \Psi^G(q+\Delta q) - \Psi^G(q)\bigr| \leq K'' \, \Delta
  q^{1-\varepsilon}
\end{equation}
for arbitrary $\varepsilon>0$, i.e., H\"older continuity of degree
$1-\varepsilon$.

In order to obtain a stronger estimate than \eqref{asbound}, one might
hope that
\begin{equation}\label{mixedvanish}
  \frac{\partial^2 \rho} {\partial q \, \partial q'} \bigg|_{q'=q} =0
  \quad \text{for all } q \in I .
\end{equation}
This would allow us to replace the exponent in \eqref{asbound} by $3/2
- \varepsilon$, which would give us in particular local Lipschitz
continuity, so that $\Psi^G$ would be differentiable
Lebesgue-almost-everywhere. But any exponent greater than 1 is too
good to be true. Indeed, as we shall show presently,
\eqref{mixedvanish} holds only for one particular density matrix
$\rho_0$, the projection to the 1-dimensional subspace of constant
functions.  For $\rho = \rho_0$, $\Psi(q)$ is constant with modulus
determined by normalization and random phase.

To see $\rho = \rho_0$, write $\rho$ in terms of its eigenfunctions
$\varphi_n(q)$ and eigenvalues $p_n$, $\rho(q,q') = \sum_n p_n \,
\varphi_n(q) \, \varphi_n^*(q')$, and observe that
\[
  \frac{\partial^2 \rho} {\partial q \, \partial q'} \bigg|_{q'=q} =
  \sum_n p_n \, \varphi_n'(q) \, \varphi_n^{\prime *}(q')
  \bigg|_{q'=q} = \sum_n p_n \, |\varphi'_n(q)|^2\,,
\]
where $\varphi_n'$ denotes the derivative of $\varphi_n$.  The only
way how this quantity can vanish for all $q$ is that all $\varphi_n$
have identically vanishing derivative and thus are constant, which
implies $\rho =\rho_0$.

\section{Concentration on the Domain of $H$}\label{sec:domain}

In this section we utilize a fourth kind of argument, different from
those of the previous sections. It is our most elegant argument,
particularly simple and direct, as it deals only with the eigenvalues
and eigenvectors, but also more abstract. This argument applies not
only to $GAP(\rho_\beta)$ but to every probability measure $\mu$ on
$\sphere(\Hilbert)$ with density matrix $\rho_\beta$.

The question we address is whether $\mu(\domain(H)) =1$. The 
answer is yes. Even more, we show that, for any \z{self-adjoint} 
$H$ \z{and almost all $\beta$ for which $Z(\beta,H) < \infty$, any 
$\mu$ with density matrix $\rho_\beta$ is supported by the domain 
of $H^\ell$ for every $\ell \in \NNN$.} (Whether $\psi \in \domain(H)$ 
implies differentiability depends of course on $H$.) \z{After that, 
we show further that
$GAP(\rho_\beta)$ for sufficiently large $\beta$ is concentrated 
on the \emph{subspace of analytic vectors} \cite{BR} of $H$.}

\begin{theorem}\label{sta:domain}
  Let $\rho$ be a density matrix on the Hilbert space $\Hilbert$,
  $\mu$ a probability measure on $\sphere(\Hilbert)$ with density
  matrix $\rho$, and $f: [0,\infty) \to \RRR$ a measurable function.
  If
  \begin{equation}\label{trrhofcond}
    \tr \bigl(\rho \, f(\rho)^2 \bigr) < \infty
  \end{equation}
  then $\mu \bigl(\domain \bigl(f(\rho) \bigr) \bigr) = 1$, where
  the domain of $f(\rho)$ can be defined, in terms of an orthonormal
  basis $\{|\varphi_n\rangle \}$ of eigenvectors of $\rho$ with
  eigenvalues $p_n$, by
  \begin{equation}
    \domain \bigl(f(\rho) \bigr) = \bigl\{ \psi \in \Hilbert: \sum_n
    \bigl| f(p_n) \sp{\varphi_n}{\psi} \bigr|^2 < \infty \bigr\}.
  \end{equation}
\end{theorem}

\begin{proof}{}
From \eqref{trrhofcond} it follows that
\[
  \EEE_\mu \sum_n \bigl| f(p_n) \sp{\varphi_n}{\psi} \bigr|^2 = \sum_n
  f(p_n)^2 \, \EEE_\mu \bigl| \sp{\varphi_n}{\psi} \bigr|^2 = \sum_n
  f(p_n)^2 \, p_n < \infty.
\]
Therefore, $\mu$-a.s.\ $\sum_n \bigl| f(p_n) \sp{\varphi_n}{\psi}
\bigr|^2 < \infty$, or $\mu \bigl(\domain(f(\rho)) \bigr) = 1$.
\end{proof}

\begin{corollary}\label{sta:Hdomain}
  Let $H$ be a self-adjoint operator on the Hilbert space $\Hilbert$.
  Suppose $Z(\beta_0,H) < \infty$ for some $\beta_0>0$, which implies
  $Z(\beta,H) < \infty$ for every $\beta > \beta_0$. Then \z{for every
  $\beta > \beta_0$ and every probability measure $\mu$ on
  $\sphere(\Hilbert)$ with density matrix} $\rho_\beta$, $\mu 
  (C^\infty(H)) = 1$, where $C^\infty (H) 
  = \bigcap_{\ell=1}^\infty \domain (H^\ell)$.
\end{corollary}

\begin{proof}{}
 For $\rho$ given by \eqref{rhobetaH}, we have that $H =
-\tfrac{1}{\beta} \log \rho + E_0 \, \mathrm{id}$ for some constant
$E_0$. Define $f(x) = (-\tfrac{1}{\beta} \log x + E_0)^\ell$ for $x>0$
and $f(x)=0$ for $x=0$.  Since $f(\rho) = H^\ell$,
Theorem~\ref{sta:domain} yields the claim if we can confirm the
condition \eqref{trrhofcond}, which we do now.

Since $\tr \, \exp(-\beta_0 H)< \infty$, there is a basis $\{
|\varphi_n\rangle : n \in \NNN\}$ of eigenvectors of $H$ with eigenvalues
$E_n$.  Furthermore, only finitely many of the eigenvalues lie below
zero, so that the set $\indexset := \{n \in \NNN: E_n > 0\}$ contains
all except finitely many numbers. Observe that for every $\beta
>\beta_0$,
\[
  \infty > \tr \, \E^{-\beta_0 H} = \sum_{n \in \NNN} \E^{-\beta_0 E_n}
  \geq \sum_{n \in \indexset} \E^{-\beta_0 E_n} > \sum_{n \in \indexset}
  \E^{-\beta E_n},
\]
and thus $Z(\beta,H) < \infty$.  For $\varepsilon>0$ with $\varepsilon
< \beta-\beta_0$, we find, for any $\ell \in \NNN$,
\[
  \infty > \sum_{n\in \indexset} \E^{-(\beta -\varepsilon) E_n} =
  \sum_{n\in \indexset} \E^{-\beta E_n} \, \E^{\varepsilon E_n} >
  \sum_{n\in \indexset} \E^{-\beta E_n} \sum_{k=0}^{2\ell}
  \frac{\varepsilon^k E_n^k}{k!} = \sum_{k=0}^{2\ell}
  \frac{\varepsilon^k}{k!} \sum_{n\in \indexset} E_n^k \, \E^{-\beta
  E_n}.
\]
In particular,
\[
  \sum_{n\in \indexset} E_n^{2\ell} \, \E^{-\beta E_n} < \infty,
\]
which implies $\tr \bigl(\rho_\beta H^{2\ell}\bigr) < \infty$.
\end{proof}

\z{If $\beta >2\beta_0$ (with $Z(\beta_0,H)< \infty$), we obtain 
the stronger result for the measure $GAP(\rho_\beta)$ 
that it is concentrated on the subspace $C^\omega(H)$ of 
analytic vectors of $H$, i.e., those vectors $\psi \in C^\infty(H)$ with}
\begin{equation}\label{analvectdef}
  \sum_{\ell=0}^\infty  \frac{\|H^\ell \psi\| \varepsilon^\ell}{\ell !} < \infty
\end{equation}
\z{for some $\varepsilon>0$ \cite{BR}. It is sufficient for $\psi 
\in C^\omega(H)$ that}
\begin{equation}\label{epsilonZabs}
  \sum_n \E^{\varepsilon |E_n|} \, \bigl| \sp{n}{\psi}\bigr| <\infty
\end{equation}
\z{because then}
\[
  \sum_{\ell=0}^\infty \frac{\varepsilon^\ell}{\ell !} \|H^\ell \psi\| = 
  \sum_{\ell=0}^\infty \frac{\varepsilon^\ell}{\ell !} \Bigl\| \sum_n E_n^\ell 
  |n\rangle \sp{n}{\psi} \Bigr\| \leq
\]
\[
  \leq \sum_n \sum_{\ell=0}^\infty \frac{\varepsilon^\ell}{\ell !} 
  |E_n|^\ell \, \bigl|
  \sp{n}{\psi} \bigr| = \sum_n \E^{\varepsilon|E_n|} \, \bigl|
  \sp{n}{\psi} \bigr| < \infty \,.
\]
\z{For $0<\varepsilon< \beta/2-\beta_0$, \eqref{epsilonZabs} is a.s.\ 
true of $\psi=\Psi^G$, and thus also of $\psi = \Psi^{GAP}$,
because, assuming without loss of generality that all $E_n>0$, 
we have by \eqref{EZabs} that}
\[
  \EEE \sum_n \E^{\varepsilon E_n} \,  \bigl| \sp{n}{\Psi^G} \bigr| =
  \sum_n \E^{\varepsilon E_n} \, \frac{\sqrt{\pi}}{2\sqrt{Z(\beta)}} \E^{-\beta E_n/2} 
  \leq \frac{\sqrt{\pi}}{2\sqrt{Z(\beta)}} \sum_n \E^{-\beta_0 E_n} < \infty\,.
\]

\section{Existence of Bohmian Velocities}

As a final remark, we mention an application of smoothness of the wave
function: differentiability is needed in Bohmian mechanics
\cite{survey}, a theory ascribing trajectories to the particles of
nonrelativistic quantum mechanics.  This is because the Bohmian law of
motion, which for $N$ particles with masses $m_1, \ldots, m_N$ at the
configuration $Q(t) = (\boldsymbol{Q}_1(t), \ldots,
\boldsymbol{Q}_N(t))$ reads
\[
  \frac{\D\boldsymbol{Q}_i}{\D t} = \boldsymbol{v}_i^\psi(Q) =
  \frac{\hbar}{m_i} \Im \frac{\psi^* \nabla_i \psi}{\psi^* \psi}(Q),
\]
involves the derivative of the wave function. Suppose the wave
function $\psi$ is chosen at random according to the canonical
distribution $GAP(\rho_\beta)$ with inverse temperature $\beta$. Then any
condition on the Hamiltonian entailing that $\psi$ is a.s.\ smooth
also implies that the Bohmian velocity vector field $v^\psi$ on
configuration space $\conf = \RRR^{3N}$, whose $i$-th component is
$\boldsymbol{v}_i^\psi$, is a.s.\ well defined everywhere outside the
nodes of $\psi$.

The analogous conclusion holds, as we shall explain presently, for the
numerous further variants of Bohmian mechanics that have been
considered (such as Bohmian mechanics on curved spaces, on the
configuration space of a variable number of particles \cite{crea2B},
for wave functions that are cross-sections of a complex vector bundle,
and variants suitable for the Dirac equation or for photons).  The
laws of motion of these variants,
\[
  \frac{\D Q}{\D t} = v^\psi(Q),
\]
are defined by giving the appropriate expression for the velocity
vector field $v^\psi$ on the manifold $\conf$, and these definitions
of $v^\psi$ can be summarized by the formula \cite{crea2B}
\begin{equation}\label{Bohm}
  v^\psi(q) \cdot \nabla f(q) = \Re \frac{\psi^*(q) \, \bigl(
  \tfrac{\I}{\hbar} [H,f] \psi\bigr)(q)} {\psi^*(q) \,
  \psi(q)}\quad \forall f \in  C_0^\infty(\conf)\,.
\end{equation}
Here, $f:\conf \to \RRR$ is an arbitrary smooth function with compact
support playing the role of a coordinate function, and numerator and
denominator involve inner products in the value space of $\psi$ (which
may be a fiber space of a vector bundle of which $\psi$ is a
cross-section).  For $\psi \in \domain(H)$ and $f \in
C_0^\infty(\conf)$, the right hand side of \eqref{Bohm} will be well
defined since multiplication by $f$ maps the domain of $H$ to itself,
\z{since} $H$ is the sum of a differential operator (of up to second
order) and a multiplication operator.  Since the $f$'s from
$C_0^\infty(\conf)$ suffice for determining $v^\psi$ (up to changes on
a null set), one obtains indeed a vector field $v^\psi$, defined on
$\conf \setminus \{q: \psi(q) =0\}$, for every $\psi$ from the domain
of $H$.

Hence, every wave function $\psi$ from the domain of $H$ is
sufficiently regular to define a Bohm-type velocity field. By
Corollary~\ref{sta:Hdomain}, the random wave function $\Psi$ with the
thermal equilibrium distribution $GAP(\rho_\beta)$ possesses a velocity
field $v^\Psi$ with probability one, provided that there is
$\beta_0<\beta$ with $Z(\beta_0, H) <\infty$.

\bigskip

\noindent \textbf{Acknowledgments.} We gratefully acknowledge that the
consideration of the magnitude of difference quotients (carried out in
Section~\ref{sec:increments}) was suggested to us by Detlef D\"urr of
LMU M\"unchen, Germany; that the result \eqref{GSresult} was
pointed out to us by Matthias Birkner of WIAS Berlin, Germany; \z{and 
that analyticity as a consequence of our estimates was pointed out to 
us by Jean Bricmont of UC Louvain-la-neuve, Belgium}. We
thank Sheldon Goldstein of Rutgers University, USA, for helpful
discussions.

\end{document}